\documentclass[fleqn,11pt]{wlscirep}
\usepackage[utf8]{inputenc}
\usepackage[T1]{fontenc}
\usepackage{hyperref}

\usepackage{array}
\usepackage{booktabs}
\usepackage{color, colortbl}
\usepackage{xcolor}
\definecolor{Gray}{gray}{0.96}
\usepackage[final]{pdfpages}
\makeatletter
\AtBeginDocument{\let\LS@rot\@undefined}
\makeatother
\title{Effects of shady environments on fish collective behavior}

\author[1,*]{Haroldo V. Ribeiro} 
\author[2,$\dagger$]{Matthew R. Acre} 
\author[2]{Jacob D. Faulkner} 
\author[1]{Leonardo R. da Cunha} 
\author[3]{Katelyn M. Lawson} 
\author[4]{James J. Wamboldt} 
\author[4]{Marybeth K. Brey} 
\author[5]{Christa M. Woodley} 
\author[2]{Robin D. Calfee} 

\affil[1]{Departamento de F\'isica, Universidade Estadual de Maring\'a, Maring\'a, PR 87020-900, Brazil}
\affil[2]{U.S. Geological Survey, Columbia Environmental Research Center, 4200 New Haven Road, Columbia, Missouri, 65201, USA}
\affil[3]{Department of Biological Sciences, Auburn University, 101 Rouse Life Sciences Building, Auburn, Alabama, 36849, USA}
\affil[4]{U.S. Geological Survey, Upper Midwest Environmental Sciences Center, 2630 Fanta Reed Road, La Crosse, Wisconsin, 54603, USA}
\affil[5]{U.S. Army Corps of Engineers, Engineer Research and Development Center, 3909 Halls Ferry Road, Vicksburg, Mississippi, 39180, USA}

\affil[*]{email: hvr@dfi.uem.br}
\affil[$\dagger$]{email: macre@usgs.gov}

\begin{abstract}
Despite significant efforts devoted to understanding the underlying complexity and emergence of collective movement in animal groups, the role of different external settings on this type of movement remains largely unexplored. Here, by combining time series analysis and complex network tools, we present an extensive investigation of the effects of shady environments on the behavior of a fish species (Silver Carp \textit{Hypophthalmichthys molitrix}) within earthen ponds. We find that shade encourages fish residence during daylight hours, but the degree of preference for shade varies substantially among trials and ponds. Silver Carp are much slower and exhibit lower persistence in their speeds when under shade than out of it during daytime and nighttime, with fish displaying the highest persistence degree and speeds at night. Furthermore, our research shows that shade affects fish schooling behavior by reducing their polarization, number of interactions among individuals, and the stability among local neighbors; however, fish keep a higher local degree of order when under shade compared to nighttime positions.
\end{abstract}

\begin{document}
\rfoot{\small\sffamily\bfseries\thepage/16}%

\flushbottom
\maketitle

\thispagestyle{empty}

\section*{Introduction}

Collective movement is a common behavior of many animal species that congregate into flocks, herds, or schools~\cite{camazine2001self, sumpter2010collective}. These collective movements yield a myriad of spatiotemporal patterns that are well-known to provide amazing visual spectacles in nature~\cite{vicsek2012collective}. Beyond a contemplative beauty, understanding the underlying complexity and emergence of collective movement in animal groups has been for a long time part of the agenda of researchers from several disciplines. Significant effort has been devoted to proposing and investigating minimal models (such as the Vicsek model~\cite{vicsek1995novel}) capable of connecting individual-level rules to the observed macroscopic phenomena~\cite{toner1995long, bialek2012statistical, gautrais2012deciphering, mateo2017effect, charlesworth2019intrinsically}. However, advances in modeling approaches have been, in large part, uncoupled from empirical observations which in turn creates a gap between theory and experiment that only recently started to decrease. Indeed, recent studies relying on new methods and tools for tracking animals in groups have shed light on empirical questions that would be considered unworkable a few years ago~\cite{heupel2006automated, straw2011multi, perez-escudero2014idtracker, walter2021trex, tuia2022perspectives, jetz2022biological}. Examples include the identification of influential neighbors of fish in moving groups~\cite{jiang2017identifying}, effects of predation on shoaling fish interactions~\cite{herbert2017predation} and on collective escape of pigeons~\cite{papadopoulou2022self}, emergence of swirling motion~\cite{nuzhin2021animals}, and the role of social interactions on developmental trajectories of honey bees~\cite{wild2021social}.

Despite the fascinating research that has already been conducted, most of these works are focused on a single external setting and much less is known about the effects of different external settings on the collective behavior of animals~\cite{schaerf2017effects}. In this context, an exciting possibility refers to the selection of overhead structures that create shade (hereafter shade structures) over non-shaded areas by fish species. Several species have shown an affinity for shade structures which may also play a role in the dispersal and establishment of invasive species~\cite{crook1999relationships}. Indeed, there is empirical evidence that shade increases residence time of some fish species. For instance, Bluegill (\textit{Lepomis macrochirus}) -- a freshwater fish native to the United States -- is anecdotally known to shelter in the shade of trees along banks and indeed displays a strong preference for shaded areas even in the presence of predators~\cite{mccartt1997light}. Shade structures are also known to attract marine fish species to mangroves habitats~\cite{cocheret2004attracts}, increase the number of salmon (\textit{Oncorhynchus kisutch}) in stream channels~\cite{mcmahon1989influence} and increase habitat use by the Three-Spined Stickleback (\textit{Gasterosteus aculeatus})~\cite{jones2019shade} and Lake Whitefish (\textit{Coregonus clupeaformis})~\cite{scherer1998preference}. The preference for shade is often associated with low predation risk, not only in fish~\cite{mccartt1997light, mcmahon1989influence}, but also in birds~\cite{fernandez2007changes} and rodents~\cite{mandelik2003structurally, carr2014wintering}. Alternatively, some species such as the Zebrafish (\textit{Danio rerio})~\cite{jones2019shade} exhibit no clear preference for shade, while others such as the Chinook Salmon (\textit{Oncorhynchus tshawytscha}) seem to perceive shade as a risky environment~\cite{sabal2021shade}. 

Considering the interest in understanding the preference for shade in several species and the lack of works focused on studying collective behavior under different contexts, here we investigate how shade affects the various facets of the collective behavior of Silver Carp (\textit{Hypophthalmichthys molitrix}). Silver Carp are native to eastern Asia and were imported to North America in the 1970s to be used in aquaculture, but soon escaped. This invasive species is now highly abundant in the Mississippi River basin and is further expanding to other North American basins with impacts on native fish populations~\cite{asian2005kolar}. Recent work has found that Silver Carp spend time near a large lock and dam on the Mississippi River~\cite{fritts2021lock} during the day, suggesting that shade may encourage use. Moreover, grouping tendencies of Silver Carp may be related to predator avoidance but remain poorly understood~\cite{ghosal2016invasive}. Understanding the behavior of Silver Carp and how it changes under different external contexts may reveal important patterns in habitat utilization with potential implications for species management. The main goal of our work is thus to understand how the collective behavior of these fish changes under different external settings related to the presence of shade structure during daylight hours (in and out of shade) and at night.

To do so, we rely on large-scale experiments in four earthen ponds where the position of small schools ($\sim$10 individuals) of Silver Carp are tracked using acoustic telemetry arrays over seven 48-hour trials. In these experiments, a shaded environment is created by anchoring a large floating platform within each pond, allowing us to quantify shade selection and, more importantly, how the behavior of these fish differ when under or out of shade structures during the day and at night. Silver Carp exhibited a preference for shade structures during daylight hours, but the intensity of this preference varies across trials and ponds. Our research shows that Silver Carp are slower under shade than out of shade; however, it is at night that these fish reach their highest speeds. Movement persistence also decreases when Silver Carp are under shade and increases at night. Furthermore, shade affects the schooling behavior of these fish by reducing their polarization, number of interactions among individuals, and the stability in local neighbors. Still, we find that Silver Carp preserve an intermediate local degree of order when under shade by keeping track of their neighbors' positions even in swarm-like states.

In what follows, we describe the experiments, the collection of fish trajectories, and an analysis of the preference for shaded areas of this species. Next, we quantify the effects of shade on fish speed and the degree of persistence in speed time series. Then, focusing on collective behavior, we estimate two order parameters related to the degree of alignment and rotation in Silver Carp schools. Finally, we create a complex network representation of fish interactions based on their spatial relations to highlight how shade affects the relationships among Silver Carp.

\section*{Results}

To investigate the effects of shade on the behavior of Silver Carp, we use the results of experiments carried out in four earthen ponds with water surface dimensions of approximately 40 meters by 25 meters and depth range from 1 to 2 meters (see Materials and Methods). These experiments comprise seven trials in each pond where fish positions are recorded using acoustic telemetry arrays over two full days. A 5-meter square floating platform is anchored in one of five positions in each pond during each trial which creates a shaded environment during daylight hours. Experiments started with 10 individuals in each pond, but this number decreased over the trials due to transmitter loss (see Materials and Methods and Table~\ref{table:1}). 

\begin{figure*}[!ht]
\begin{center}
\includegraphics[width=1.\linewidth,keepaspectratio]{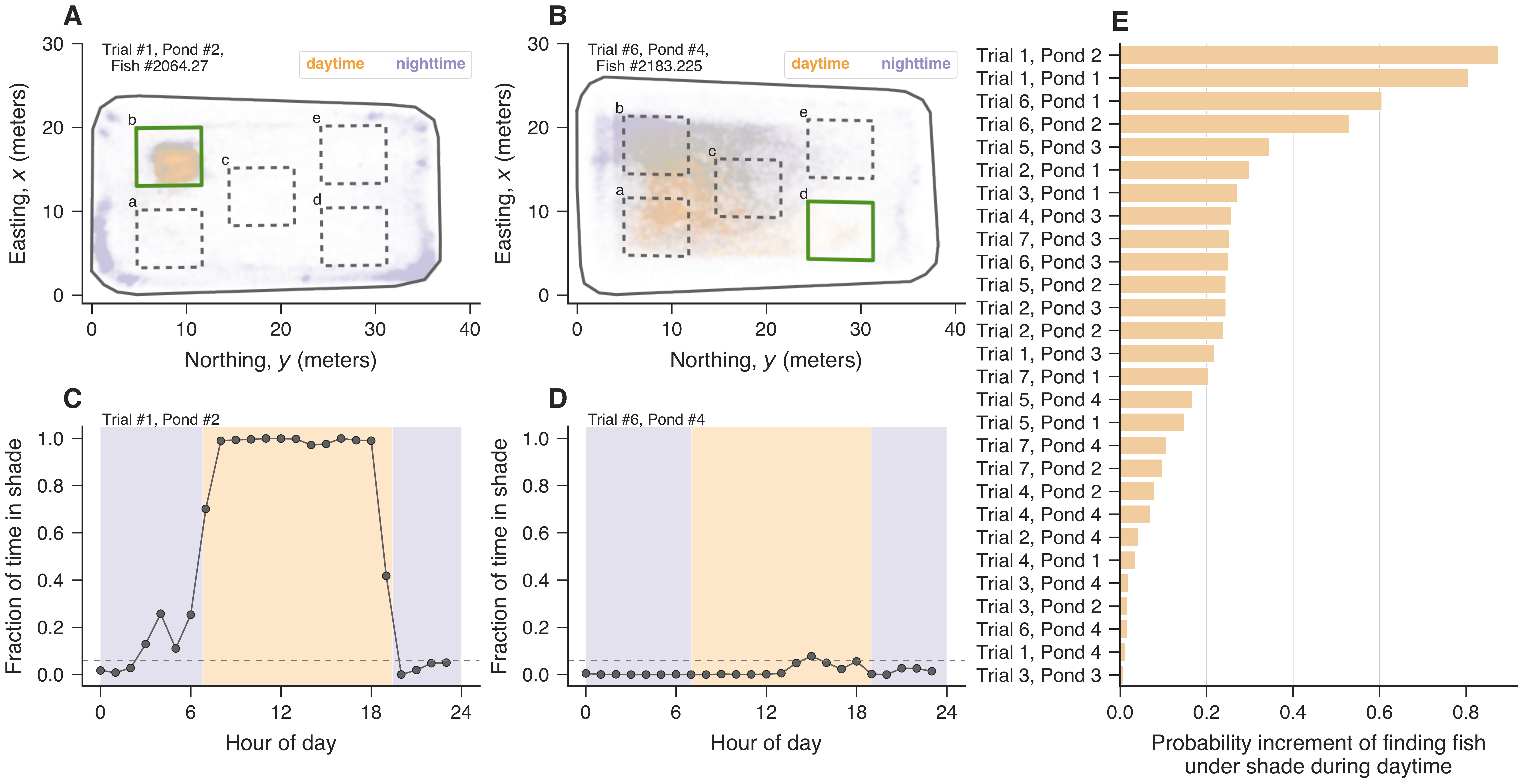}
\end{center}
\caption{Silver Carp prefer shade structures during daylight hours. (A)-(B) Density maps of two fish positions from two different trials and ponds. Orange hues indicate positions recorded during daylight hours, while purple hues represent positions collected at night. The gray solid lines represent the pond limits and the thick green squares indicate the actual location of the shade structure in the particular trials and ponds shown in the top left of the panels. The dashed squares indicated by lowercase letters (from `a' to `e') represent all possible shade positions. We observe that the fish in panel (A) has an extreme preference for the shaded area during daylight hours, while the same preference is absent for the fish in panel (B). (C)-(D) Differences in shade selection as a function of the hour of day. The circles show the fraction of all fish positions recorded under the shade structures within a one-hour time window for two trials and ponds. The dashed line represents the expected fraction of positions under shade if fish would move randomly over the ponds, and the background colors indicate daytime (orange) and nighttime (purple). Fish from pond \#2 remained under shade during nearly all daylight hours along the two observation days in trial \#1 and avoided shade at night. Fish from pond \#4 selected the shade structure much more rarely during daylight hours (particularly in the mornings) in trial \#6. (E) Probability increment of finding a position under a shade structure during daylight hours, as estimated via logistic regression for each trial and pond. The relationship between a fish selecting shade or not and the binary independent variable of day or night varies substantially but is nonetheless significant for all trials and ponds ($p$-values~$<0.001$).}
\label{figure1}
\end{figure*}

Figures~\ref{figure1}A and \ref{figure1}B show density maps for the position of two fish from different ponds and trials during daytime and nighttime (see Fig.~S1 for all ponds and trials). We observe that one of these fish shows strong preference for the shade structure over the rest of the pond during daylight hours, while the other fish does not seem to have the same preference. Similar variability occurs when we consider the position of all fish and evaluate the shade selection by calculating the fraction of time they spend under shade structures over the hours of day. Figures~\ref{figure1}C and \ref{figure1}D exhibit this analysis for all fish from the same ponds and trials used to create the density maps. We note that the individual behaviors of fish in Figs.~\ref{figure1}A and \ref{figure1}B agree with the fractions of time spent under shade structures by their schools. Fish from the pond and trial of Fig.~\ref{figure1}C spend most of their time under shade during daytime and ignore the shade structure at night. Conversely, fish from pond and trial depicted in Fig.~\ref{figure1}D never select the shade structure between midnight and noon, and were rarely under shade in the afternoon and early evening hours. These two ponds and trials are indeed extreme examples of variability in shade preference (see Fig.~S2 for all ponds and trials). We quantify the preference for shade structures during daytime using a logistic regression where a binary variable indicating whether day or night is used to predict whether fish is under shade or not (see Materials and Methods). We find that the association between these covariates substantially varies among ponds and trials, but daylight significantly increases the probability of finding fish under shade structure in all cases (Fig.~\ref{figure1}E). To account for potential location bias within each pond, we have performed a control-like analysis in which the shade preference in a given trial and pond is estimated using the shade location from the next and previous trials (that is, not the actual shade location). We find that the probability increment of finding fish under shade during daytime vanishes in both control-like analyses (Fig.~S3), indicating that the association between fish selecting shade or not and the binary independent variable of day or night cannot be explained by possible biased preferences induced by the ponds themselves. Moreover, there is no recognizable pattern in shade selection across ponds and trials (Figs.~S1 and S2) and we find no correlation between the fraction of time spent under shade over the hours of day and temperature or light intensity throughout the trials (Figs.~S4, S5, and S6).

\begin{figure*}[!ht]
\begin{center}
\includegraphics[width=0.8\textwidth,keepaspectratio]{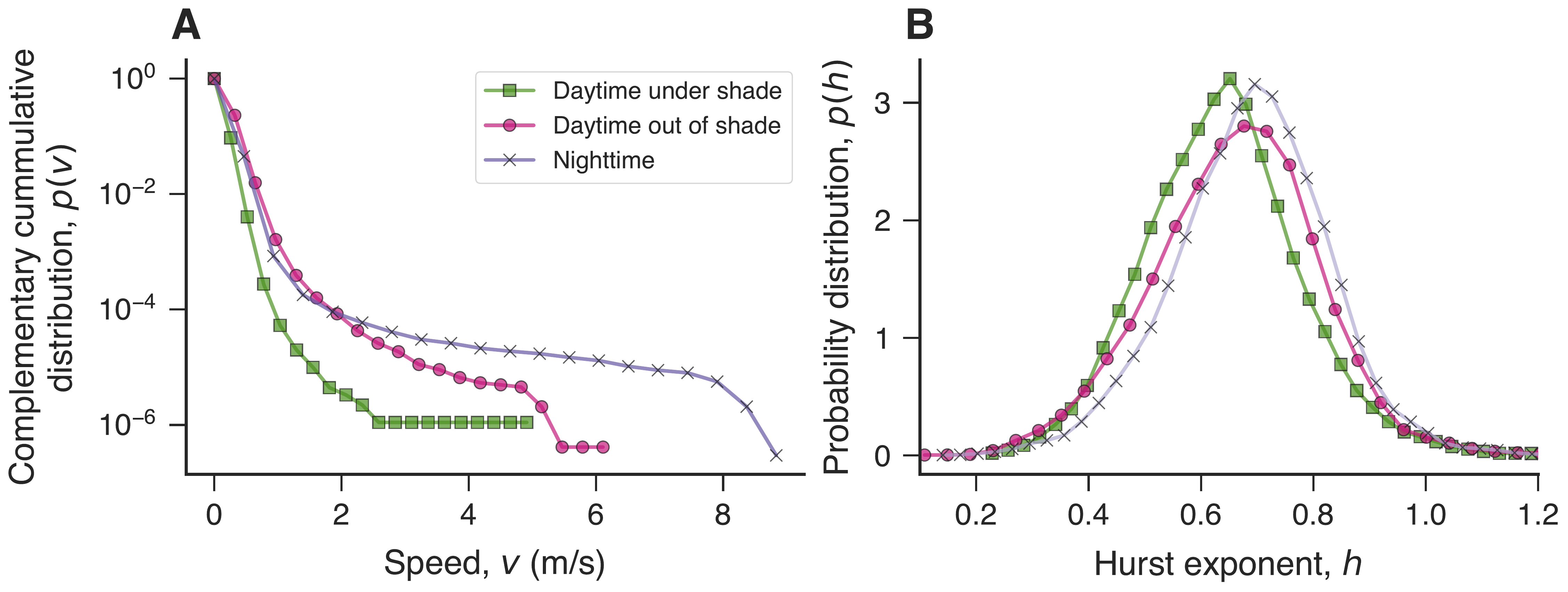}
\end{center}
\caption{Fish speeds differ during daylight hours in or out of shade structures and at night. (A) Complementary cumulative distributions of fish speed $v$ grouped into three categories: daylight hours under shade (green squares), daylight hours out of shade (pink circles), and nighttime positions (purple crosses). Fish move slower under shade structures ($\langle v \rangle = 0.1286 \pm 0.0001$ m/s; mean $\pm$ standard error of the mean, unless otherwise stated) than out of them ($p$-value~$< 0.001$, permutation test). The average speed is higher when fish are out of shade during daylight hours ($\langle v \rangle = 0.2324 \pm 0.0001$ m/s) than at night ($\langle v \rangle = 0.2086 \pm 0.0001$ m/s; $p$-value~$< 0.001$, permutation test). However, the speed distribution is much more skewed at night [skewness $4.1 \pm 0.3$ vs. $1.8 \pm 0.1$ (under shade) and $1.7 \pm 0.1$ (out of shade); $p$-values~$< 0.001$, permutation tests], indicating that fish reach their highest speeds during nighttime. (B) Probability distribution of the Hurst exponent $h$ obtained via kernel density estimation from speed time series grouped into the same three categories as the previous panel. Speed time series display long-range persistent behavior in the three categories, but the average Hurst exponent is slightly smaller when fish are under shade during daylight hours ($\langle h \rangle = 0.639 \pm 0.002$) than when they are out of shade during daytime ($\langle h \rangle = 0.661 \pm 0.002$; $p$-value~$< 0.001$, permutation test) or at night ($\langle h \rangle = 0.689 \pm 0.002$; $p$-value~$< 0.001$, permutation test).}
\label{figure2}
\end{figure*}

While there is evidence supporting the hypothesis that shade structure encourages Silver Carp residence, we do not know whether the behavior of these fish substantially differs when they select shade structures compared to when they are out of shade during daytime or nighttime. And to start addressing this question, we calculate the speed of fish, grouping data into three categories (see Materials and Methods): fish under shade during daytime, fish out of shade during daytime, and fish at night. We have not categorized night positions by location because the interaction with the shade structure was minimal at night. Figure~\ref{figure2}A depicts the complementary cumulative distributions of fish speed (that is, the probability $p(v)$ of finding a fish with speed larger than $v$) grouped into these three categories. These results clearly show that fish move much slower under shade structures during daytime, averaging $\langle v \rangle \approx 0.13$ m/s, with fish rarely exceeding one meter per second. On average, fish are almost twice as fast ($\langle v \rangle \approx 0.23$ m/s) when out of shade during daylight hours. Fish are slightly slower at night ($\langle v \rangle \approx 0.21$ m/s) than out of shade during daytime, but they reach the highest speeds recorded in our trials at night. These differences in speed suggest that fish may use these structures to rest and shelter. Conversely, the higher speeds out of shade are likely to be associated with foraging and feeding behaviors, as Silver Carp are pelagic filter feeders~\cite{spataru1985feeding, lazzaro1987review, voros1997size}. These high speeds during nighttime also suggest that these fish use the cover of dark to safely explore the ponds as this visual condition may decrease predation risks in real environments.

In addition to the speed probability distribution, dynamical features of fish speed time series may also indicate different behaviors under or out of shade during daytime or nighttime. An interesting possibility is to quantify the degree of persistence in fish speed. If fish are idler under shade and move more objectively out of it, we expect speed time series to be less persistent when fish are under shade during daytime. To verify this hypothesis, we estimate the Hurst exponent $h$ from speed time series (see Materials and Methods) grouped into the same categories used above. Hurst exponents $h>0.5$ indicate long-range persistent behavior such that time series increments are more likely to be followed by increments with the same signal, and the closer $h$ is to 1, the higher the persistence degree. Values of $h<0.5$ represent an anti-persistent behavior in which time series increments alternate signs more likely than by chance. A time series is not correlated if $h=0.5$. Figure~\ref{figure2}B shows that fish speed is marked by long-range persistence under or out of shade during daytime and nighttime (about 90\% of the time series have $h>0.5$). There are nonetheless appreciable differences in the distribution of $h$ among the three situations. We observe that the Hurst exponent distribution is slightly shifted to the left for positions under shade during daylight hours when compared with the other two situations, which in turn makes the average value of $h$ under shade ($\langle h \rangle = 0.639 \pm 0.002$; mean $\pm$ standard error of the mean, unless otherwise stated) smaller than out of shade during daytime ($\langle h \rangle = 0.661 \pm 0.002$) and at night ($\langle h \rangle = 0.689 \pm 0.002$). These results support the hypothesis that fish move less persistently under shade and also indicate that their movements are more persistent at night.

\begin{figure*}[!ht]
\begin{center}
\includegraphics[width=1.\textwidth,keepaspectratio]{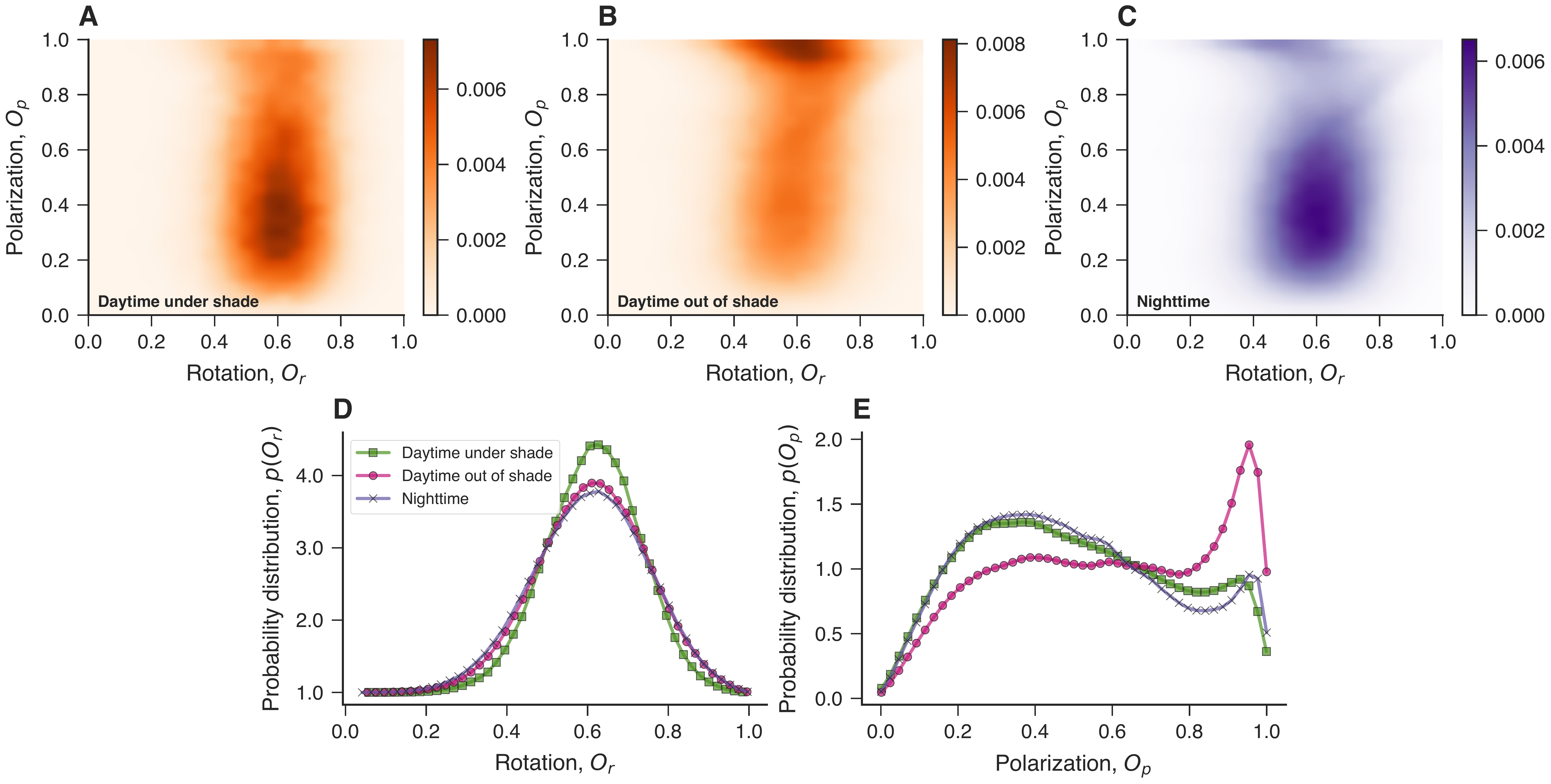}
\end{center}
\caption{Shade affects schooling behavior of Silver Carp. Joint probability distributions of rotation ($O_r$) and polarization ($O_p$) order parameters when grouping fish positions into three categories: (A) daylight hours under shade, (B) daylight hours out of shade, and (C) nighttime positions. Marginal distributions of the (D) rotation and (E) polarization order parameters for the same three categories obtained via kernel density estimation. Apart from the slightly more localized distribution of $O_r$ during daylight hours under shade, there is no obvious difference in the degree of collective rotation among the three categories. Conversely, the degree of alignment measured by $O_p$ is greater when fish are out of shade during daylight hours than in the other two categories.}
\label{figure3}
\end{figure*}

Another intriguing question is whether shady environments affect the collective behavior of Silver Carp. To investigate this possibility, we calculate two order parameters that quantify the degree of alignment and the degree of rotation in schools of fish~\cite{kolpas2007coarse, tunstrom2013collective}. As defined in Materials and Methods, these are the polarization order parameter $O_p$ and rotation order parameter $O_r$. The values of $O_p$ range from $0$ to $1$ and measure how aligned the collective movement of fish is: $O_p\approx0$ indicates a swarm-like state where fish move disorderly in different directions, while $O_p\approx1$ represents a polarized-like state in which fish move aligned with each other. The values of $O_r$ are also constrained between $0$ and $1$, and this order parameter describes the school's degree of rotation around its center of mass: $O_r\approx0$ represents a state with no rotation while $O_r\approx1$ indicates a strong rotating or a milling state. We calculate the values of $O_p$ and $O_r$ for each instant of time in all ponds and trials with five or more fish and group these values in the same three categories used in previous analyses (positions under or out of shade during daytime and positions at night). Figures~\ref{figure3}A, \ref{figure3}B and \ref{figure3}C depict the joint probability distribution of the two order parameters for each category, whereas Figs.~\ref{figure3}D and \ref{figure3}E show the corresponding marginal distributions for the rotation and polarization order parameters, respectively.

The joint probability distributions of $O_p$ versus $O_r$ show that the degree of alignment of these fish differs among the three categories, with schools of fish exhibiting higher $O_p$ values when out of shade during the daytime (Fig.~\ref{figure3}B) than in the other two categories. These distinct behaviors become even more clear when noticing that the marginal distribution of $O_p$ has a peak close to 1 for positions out of shade during daytime (Fig.~\ref{figure3}E). The average value of this order parameter is higher when fish are outside of the shade structure ($\langle O_p\rangle \approx0.59$) than under shade during daytime ($\langle O_p\rangle \approx 0.50$) and at night ($\langle O_p \rangle\approx 0.50$); furthermore, the probability of finding individuals in strongly aligned states ($O_p>0.9$) is about two times greater outside of shade structures than in the other two categories. These results thus suggest that fish may adopt highly polarized states as a strategy for evading possible predators when out of shade during daytime. Intriguingly, the degree of polarization under shade during daylight hours (Fig.~\ref{figure3}A) is similar to the one observed at night (Fig.~\ref{figure3}C). Thus even though fish move faster and more persistently during nighttime than under shade structures during daytime, the degree of alignment among group members is low and does not differ much between these two situations. These low polarized states during nighttime and under shade during daytime could represent behavioral responses to reduced or low predation pressure at night and under shade structures during the day.

Conversely, the joint probability distributions of the order parameters shown in Fig.~\ref{figure3} do not indicate much difference in the degree of rotation among the three categories. The marginal distributions of $O_r$ (Fig.~\ref{figure3}D) show that the degree of rotation is only slightly less spread under shade during daytime, but the average value of this order parameter is approximately the same ($\langle O_r\rangle \approx 0.61$) in the three situations. Fish rarely adopt a strong rotating group state in our experiments, such that only $\approx$1\% of timestamps are characterized by $O_r>0.9$ regardless of the position category. 

\begin{figure*}[!ht]
\begin{center}
\includegraphics[width=1.\textwidth,keepaspectratio]{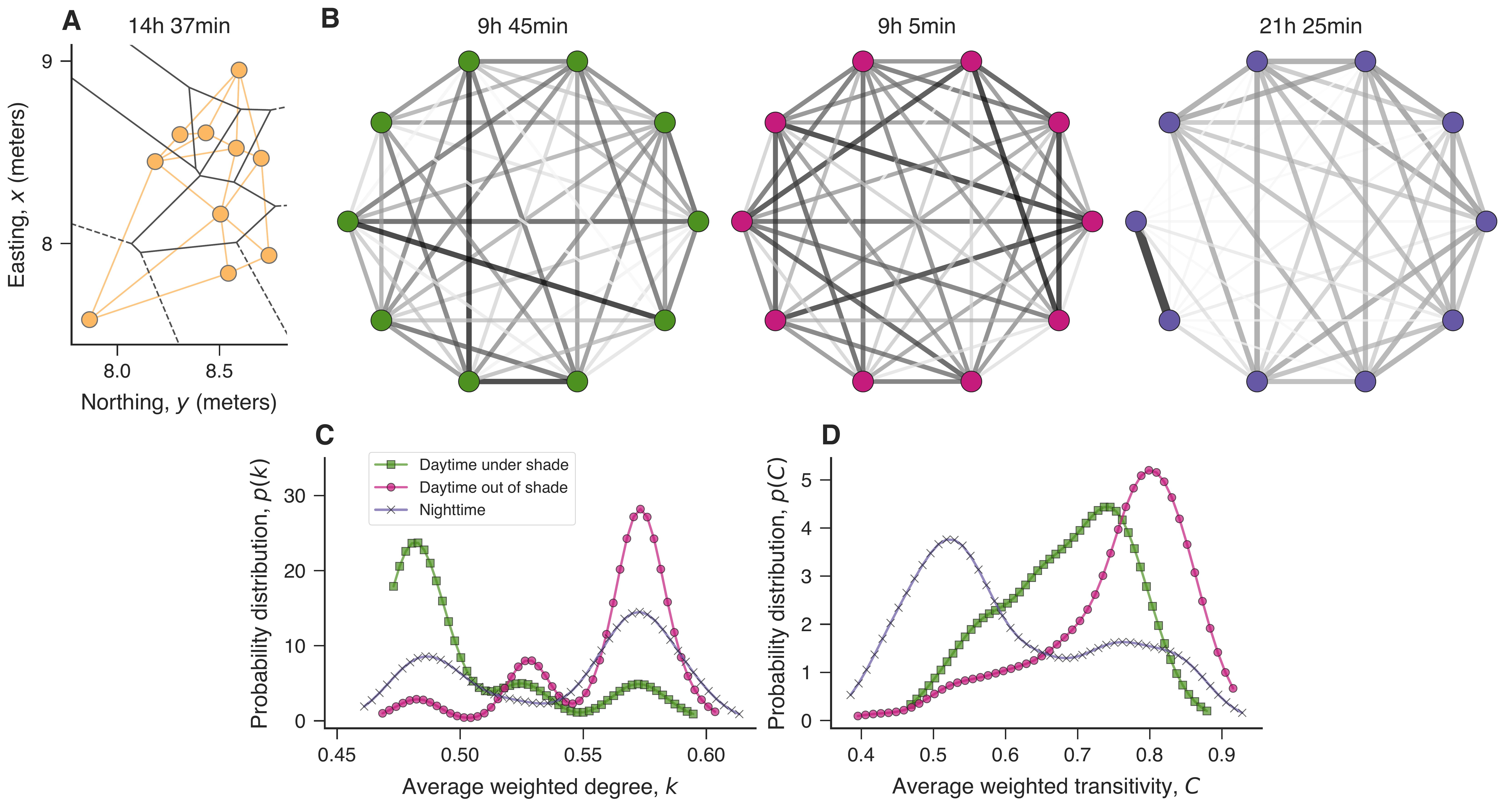}
\end{center}
\caption{Interaction networks of Silver Carp estimated during daylight hours under or out of shade structures and at night. (A) Example of Voronoi tessellation of a group of ten fish. Black lines delineate the Voronoi cells, while the orange circles indicate the fish positions for this particular time (14h 37min). The interaction network is created by connecting individuals sharing a boundary in the Voronoi diagram (orange lines show all links in this example). (B) Examples of weighted networks obtained by aggregating the interaction networks within 10-minute intervals. The edge weights represent the fraction of time a pair of fish remained neighbors in the Voronoi diagram during the time interval divided by the number of fish minus one. The darker and wider the line between individual nodes, the stronger the connection between those fish. Probability distribution of the (C) average weighted degree $k$ and (D) average weighted transitivity $C$ when grouping data into three categories: daylight hours under shade (green squares), daylight hours out of shade (pink circles), and nighttime positions (purple crosses).}
\label{figure4}
\end{figure*}

To further investigate the effects of shade on the collective behavior of Silver Carp, we create a complex network representation of fish interactions based on their spatial relations. As detailed in Materials and Methods, we first obtain the Voronoi tessellation related to fish positions in a given time and draw connections between individuals sharing a boundary in the Voronoi diagram (Fig.~\ref{figure4}A). This type of spatial relationship between individuals approximates the properties of sensor networks and is thus a good proxy for fish interactions~\cite{strandburg2013visual}. Next, we aggregate information about all connections within a 10-minute window to create a weighted network where fish are the nodes, and the weighted connections among them represent the fraction of time a pair of fish remained neighbors during the time interval divided by the number of fish minus one. Finally, we group these networks into the same three categories used previously: daylight hours under shade, daylight hours out of shade, and nighttime positions. Figure~\ref{figure4}B depicts a visualization of typical networks for each of these categories. In these visualizations, connections between individuals that remained neighbors in the Voronoi diagram for a considerable fraction of the time are represented by wide and dark edges, while thin and light gray edges indicate pairs of fish that were neighbors much less frequently.

A simple visual inspection of the networks in Fig.~\ref{figure4}B indicates that the interactions among fish during daylight hours under or out of shade and at night are different. The network representing positions out of shade during daytime has more intense connections among several individuals, while the other networks display a smaller number of such connections (particularly during nighttime). To systematically explore these differences, we first evaluate the average weighted degree $k$ of these networks (see Materials and Methods). The weighted degree $\kappa$ is the sum of weights of edges connected to a given node, and therefore it represents the sum of the fractions of time of all fish that were neighbors of a given fish divided by the total number of fish minus one. This quantity is thus an indicator of the average number of neighbors (weighted by the fraction of time and normalized by the number of fish) a fish had during the 10-minute interval; for instance, if fish remain in a star-like configuration, the central fish would have a value of $\kappa$ approximately equal to one, while $\kappa$ is approximately equal to one over the number of fish minus one for all peripheral fish. The average weighted degree $k$ is the average value of the weighted degree $\kappa$ over all network nodes, and it is thus a proxy for the number of interactions among fish in the school.

Figure~\ref{figure4}C shows the probability distribution of the average weighted degree $k$ when grouping the networks into the three categories. We observe a clear distinction between networks related to under and out of shade positions during daylight hours --- positions under shade yield networks with lower average weighted degrees than positions out of shade. This result shows that fish interact less among themselves under shade, and because we know polarization is low under shade (Fig.~\ref{figure3}E), we can further assert that this lower number of interactions is partially explained by the lack of stable formations that characterize swarm-like states. Conversely, the high number of interactions displayed by fish out of shade during daytime reflects more stable and polarized formations in which fish keep track of their neighbors' directions. In its turn, the distribution of the average weighted degree associated with nighttime positions is more uniform than the two previous cases, indicating that the number of interactions at night is more random.

We have also calculated the average weighted transitivity coefficient $C$~\cite{onnela2005intensity} (see Materials and Methods) of these networks under the three conditions. For unweighted networks, the transitivity coefficient (also known as the clustering coefficient) measures the density of closed triplets, indicating thus the tendency of nearest neighbors of a node to be connected~\cite{newman2010networks}. Similarly, the weighted transitivity quantifies the density of closed triplets taking into account the geometric mean of edge weights of the triangle motifs. Thus, closed triplets in which one of the edges has very low intensity contribute little to the weighted transitivity, while closed triplets having all edges with high weights contribute more. We can interpret this transitivity coefficient as an indicator of local order in fish position, such that stable local neighborhoods (that is, movements that tend to keep the same first neighbors) are expected to yield higher values of $C$ than unstable local neighborhoods.

Figure~\ref{figure4}D shows the probability distribution of the average weighted transitivity coefficient $C$ when grouping networks into the three categories. We note that the distributions of $C$ are remarkably different between networks related to nighttime and daytime positions under or out of shade. Nighttime positions yield the smallest transitivity values and a distribution that peaks around $C\approx0.5$, while daytime positions create networks with higher transitivity values. When comparing daytime positions under and out of shade, we further observe that positions out of shade produce larger transitivity values (distribution peaks around $C\approx0.85$) than positions under shade (distribution peaks around $C\approx0.75$). Daytime positions out of shade are thus marked by strong closed triad relationships, indicating that not only is the school's global movement polarized but also that the local neighborhood of fish remains very stable during movement under this condition. Conversely, and despite the similar polarization degree of movement during daytime under shade and during nighttime (Fig.~\ref{figure3}E), our results show that the formation of closed and strong triads is much less common in the latter condition. Thus, although fish often remain in swarm-like states under shade during the day and at night, positions under shade preserve an intermediate local degree of order in which, despite the different directions of movement, fish still keep track of their neighbors' positions.

\section*{Discussion}

We have presented an extensive investigation of the effects of shady environments on the behavior of Silver Carp. Similar to other fish species~\cite{mccartt1997light, cocheret2004attracts, mcmahon1989influence, jones2019shade, scherer1998preference}, we have first verified that shade encourages residence of Silver Carp during daylight hours but that the degree of preference for shade structures varies significantly among our experiments. Changes in light levels due to weather and/or water clarity may have a role in these differences as fish may use shade or shelter at greater light levels~\cite{mccartt1997light, mcmahon1989influence}. Cloud cover, sediment, or phytoplankton productivity may influence water clarity over short and long temporal scales. However, albeit crude, we have evaluated the relationship between light levels and shade use among ponds and trials, finding no strong correlation. Furthermore, water temperature was not found to correlate well with shade use. While study treatments designed to test the effect of environmental conditions would be best, we believe light intensity and water temperature played a minor role in our study. In regard to light, zooplankton and phytoplankton communities within pond increased across trials, creating reduced water clarity. However this did not result in a decreasing trend in the use of shade structure over trials. Alternatively, though our trials generated a large data set, our findings are based on four replicated ponds which may not be sufficient to derive clear patterns among these variables. 

Still, the lack of apparent environmental factors explaining the variation in shade selection allows us to hypothesize that individual differences and preferences may play a role in this matter. Nile tilapia (\textit{Oreochromis niloticus}), though taxonomically divergent, have been found to present individual variations in preferences related to the selection of different colors~\cite{maia2016history} and different substrate sizes~\cite{maia2018individuality}. Indeed, several studies evaluating animal preferences have found their choices to differ in subsequent tests~\cite{godin1995variability, johnsson2000habitat, shields2004dustbathing, gomez2005influence, browne2010consistency}. Under the hypothesis that individuality matters for shade selection, how a single individual may affect the entire group's decision of selecting the shade structure becomes a fascinating question that future research may address. Another possible explanation for the variability in shade selection is domestication, which among other changes, may yield risk tolerant individuals~\cite{saraiva2018domestication}. Fish used in our study were spawned and reared on-site and may have suppressed natural instincts, which in turn may also explain the variations in shade selection given the association between shade preference and predation risk~\cite{mccartt1997light, mcmahon1989influence, fernandez2007changes, mandelik2003structurally, carr2014wintering}. In this regard, future studies can verify whether shade selection and behavior differ between domestic and wild groups of individuals and even how these patterns would change in their natural habitats.

Beyond finding that fish are more likely to be under shade structures during the day than at night, our research reveals that the behavior of Silver Carp is quite different depending on whether fish are under or out of shade during daytime or nighttime. Fish swim much slower under shade than out of shade during the day, but it is at night that they reach their highest speeds. Moreover, speed time series display different degrees of long-range persistence among these three conditions. Long-range persistence decreases when fish are under shade and becomes more intense at night, an indication that movement is more directed at night and when not under shade.

We have also investigated how the collective behavior of this fish species changes among positions under or out of shade during the day and nighttime positions via two approaches. In the first one, we have relied on two order parameters that quantify the school's degree of polarization and rotation. We have found that Silver Carp rarely adopt strong rotating states in our experiments and that shade or nighttime do not affect this behavior. The lack of these strong rotating states may be attributed to the number of individuals used in our study. In research with Golden Shiner (\textit{Notemigonus crysoleucas}), Tunstr{\o}m \textit{et al.}~\cite{tunstrom2013collective} report that groups of 30 fish in a 2.1 meters by 1.2 meters shallow tank rarely adopt rotating group states, but as group size increases, the frequency of these states also increases. Thus, future research with greater numbers of Silver Carp may help us understand if these fish exhibit strong rotating states and whether shady environments play a role in emergence of these states. Conversely, fish assume swarm-like states with no preferential movement direction when under shade structures and adopt strong polarized states in which they move aligned with each other when out of shade during daylight hours.

In the second approach, we have created interaction networks for Silver Carp based on spatial relations obtained from a Voronoi tessellation of fish positions --- a technique that closely approximates fish sensory networks~\cite{strandburg2013visual}. We have found these interaction networks to be remarkably different for positions under or out of shade structures during daytime and nighttime positions. Fish interact less among themselves under shade and tend to keep a higher number of stable neighbors when out of shade during daytime; at night, the weighted number of interactions is more evenly distributed. We have further observed a hierarchy regarding the emergence of strong and closed triadic relationships among fish. The density of strong closed triads is much larger when fish are out of shade during daytime, decreases when under shade, and further diminishes at night.

Taken together, our findings indicate that Silver Carp decrease their speed, the persistence of their movement, and their group polarization under shade structures, which is consistent with resting or a latent state of movement. Still, these fish appear to keep some level of local order as indicated by the intermediate transitivity values in their interaction networks under this condition. During daylight hours and out of shade, Silver Carp move fast, persistently, and in group states displaying intense polarization and stable formations marked by very high transitivity values and high number of stable neighbors. These results thus suggest that Silver Carp keep a collective attention state when moving about the pond during daylight hours, presumably to forage~\cite{spataru1985feeding, lazzaro1987review, voros1997size}. Finally, these fish reach their highest speed and long-range persistence at night, whereas group polarization is similar to the under shade situation. Furthermore, interactions among fish are much weaker at night compared to the other two conditions as the interaction networks are characterized by a more uniform weighed degree distribution and low transitivity values. Thus, fish seem to take advantage of the dark to safely explore the environment in group states marked by small coordination, perhaps due to the reduced visual condition and presumed lower risk of predation. We believe these findings may have implications for managing this invasive species. For example, pairing this knowledge with the prioritization and design of non-structural barriers could lead to increased performance of these migratory deterrents. Lastly, our findings suggest that fish stay in more latent states under shade than out, providing fish with this choice of being under shade or not could also improve animal welfare~\cite{dawkins2006through}.

\section*{Materials and Methods}

\subsection*{Experiments and data extraction}

The experiments and data set used in this study comprise of seven trials that took place in four earthen ponds at the U.S. Geological Survey, Columbia Environmental Research Center, Columbia, Missouri, USA. Ponds are approximately $40$ meters long by $25$ meters in width and depth range of $1$-$2$ meters and were setup to be replicates; every attempt was made to control variation among ponds including water flow. Experiments started with ten juvenile Silver Carp ($345.2 \pm 17.5$ mm total length and $377 \pm 64.7$ g weight; mean $\pm$ standard deviation), spawned and reared on-site, placed in each pond. We have used juveniles because the husbandry of larger individuals is much more complex and also because they are expected to be sexually immature (controlling for potentially confounding behavioral variables related to different sexes). All individuals were fitted with acoustic transmitters ($20$ $\times$ $6.8$ mm with a weight of $1.1$ g, 795-LD; Innovasea Systems Inc., Boston, MA; formerly Hydroacoustic Technology Inc.) either surgically implanted into the coelomic cavity (20 individuals) or fixed to the fish externally (20 individuals). Animal work was in accordance with ARRIVE guidelines and all applicable standards for the ethics of experimentation and research integrity and approved by the U.S. Geological Survey Institutional Animal Care and Use Committee at the Columbia Environmental Research Center (IACUC number: AEH-18-CERC-01). An acoustic telemetry system (M290; Innovasea Systems Inc.) in each pond was used to obtain two-dimensional fish positions on average every 1.3 seconds, with fewer than 0.2 percent of relocations occurring at intervals greater than 10 seconds and mean spatial accuracy of 0.38 m. In addition, temperature and relative light level were monitored using a data logger (UA-002-64; HOBO Pendant) deployed underwater on the south end of each pond. A 5-meter square floating platform was placed in one of five possible positions in each pond and the vacant positions are indistinguishable from the remainder of the ponds. The dashed squares indicated by lowercase letters (from `a' to `e') in Figs.~\ref{figure1}A, \ref{figure1}B, and Fig.~S1 show the five possible positions for the shade structure in each pond. The shade structures were respectively located at positions `a', `b', `c', and `d' of ponds \#1, \#2, \#3, and \#4 in trial \#1. Fish were allowed to acclimate to the shade position for 24 hours at which time the 48 hour trial would begin (starting at 0h 00min and ending at 23h 59min of the following day). Floating platforms were sequentially moved to a new position for each trial; for instance, the locations of the shade structure over the trials in pond \#2 were: `c' $\to$ `d' $\to$ `e' $\to$ `a' $\to$ `b' $\to$ `c' $\to$ `d'. The acclimation period was repeated in between each trial. This platform creates a shady environment below it that mimics aquatic habitat that may be found near solid vertical or overhanging structures.

We have filtered out all positions from expelled transmitters in each trial by examining the position variability (standard deviation of consecutive position increments) of each fish in a one-hour sliding window. Trajectories with position variability smaller than 20\% of the overall variability of each fish were considered as potential transmitter loss and were visually inspected. All trajectories identified as expelled transmitters were entirely removed from our analysis. We have visually inspected all trajectories and filtered out time series associated with poor detection events that lead to very large (more than one hour) gaps in position detection. These gaps are attributed to anomalies in individual tag signals coinciding with restrictive settings in the transmitter data processing steps, which lead to the exclusion of tag transmissions. The final data set comprises more than 24 million positions from individuals that retained their acoustic transmitter throughout individual trials (18 of 40 individuals contributed data for all seven trials) as detailed in Table~\ref{table:1}. At the end of the seventh trial the ponds were drained and all 40 fish were recovered alive.

\begin{table}[!ht]
\renewcommand{\arraystretch}{1.25}
\centering
\caption{Data set summary. Number of position detections in each trial and pond of our data set.}\label{table:1}
\begin{tabular}{rrrr}
\toprule
 Trial &  Pond &  Number of fish &    Number of detections \\
\midrule \rowcolor{Gray}
     1 &    1 &       10 &  1,296,149 \\ 
     2 &    1 &       10 &  1,280,127 \\\rowcolor{Gray}
     3 &    1 &        8 &  1,056,631 \\
     4 &    1 &        8 &  1,068,860 \\\rowcolor{Gray}
     5 &    1 &        7 &    939,736 \\
     6 &    1 &        4 &    545,645 \\\rowcolor{Gray}
     7 &    1 &        4 &    541,051 \\
     1 &    2 &       10 &  1,257,650 \\\rowcolor{Gray}
     2 &    2 &        8 &  1,047,315 \\
     3 &    2 &        7 &    953,318 \\\rowcolor{Gray}
     4 &    2 &        7 &    883,221 \\
     5 &    2 &        8 &  1,044,909 \\\rowcolor{Gray}
     6 &    2 &        7 &    973,654 \\
     7 &    2 &        7 &    970,958 \\\rowcolor{Gray}
     1 &    3 &        9 &  1,161,989 \\
     2 &    3 &        8 &    933,214 \\\rowcolor{Gray}
     3 &    3 &        6 &    799,124 \\
     4 &    3 &        5 &    688,476 \\\rowcolor{Gray}
     5 &    3 &        5 &    713,752 \\
     6 &    3 &        5 &    720,318 \\\rowcolor{Gray}
     7 &    3 &        5 &    708,362 \\
     1 &    4 &        7 &    940,619 \\\rowcolor{Gray}
     2 &    4 &        8 &  1,123,747 \\
     3 &    4 &        7 &    940,349 \\\rowcolor{Gray}
     4 &    4 &        5 &    659,301 \\
     5 &    4 &        3 &    420,820 \\\rowcolor{Gray}
     6 &    4 &        3 &    419,256 \\
     7 &    4 &        3 &    415,454 \\\rowcolor{Gray}
\bottomrule
\end{tabular}
\end{table}

\subsection*{Speed time series}

To estimate the speed time series of fish, we first segment fish trajectories into two classes, representing positions in and out of shade structures. Within those classes, we then select trajectory segments with more than 50 consecutive observations and calculate speed using distance and time between consecutive positions. Finally, these speed time series were classified into three groups: daylight hours under shade, daylight hours out of shade, and nighttime positions. 

\subsection*{Detrended fluctuation analysis}

We have estimated the Hurst exponent $h$ from the speed time series segments $v(t)$ via detrended fluctuation analysis (DFA)~\cite{peng1994mosaic,kantelhardt2001detecting}. DFA consists of the following steps. We start by defining the integrated profile series $Y(t) = \sum_{i=1}^t [v(i) - \langle v(t) \rangle]$, where $\langle v(t) \rangle$ is the average speed of the time series segment. Next, we sample $Y(t)$ into $m_n=m/n$ non-overlapping partitions of size $n$, where $m$ is the length of the time series. For each partition, we adjust a linear model and subtract it from $Y(t)$, defining the detrended profile at the scale $n$ as $Y_n(t) = Y(t) - p_j(t)$, where $p_j(t)$ represents the adjusted linear function in the $j$-th partition. Finally, we calculate the root-mean-square fluctuation function $F(n) = \sum_{j=1}^{m_n} \langle Y_n(t)^2\rangle_j/m_n$, where $\langle Y_n(t)^2\rangle_j$ is the mean-square value of $Y_n(t)$ over data in the $j$-th partition. For self-similar time series, $F(n)$ displays a power-law dependence on $n$, $F(n)\sim n^h$, where $h$ is the Hurst exponent. We estimate $h$ by calculating the slope of the linearized version of the power-law relation ($\log F(n) \sim h \log n$) via ordinary least squares method. Values of $h>0.5$ indicate long-range persistent behavior, while $h\approx0.5$ represents uncorrelated or short-range correlated time series. We have also determined that the Hurst exponents from shuffled versions of the speed time series are around $0.5$, confirming the existence of long-range correlations.
 
\subsection*{Order parameters}

To calculate the order parameters, we have first interpolated the trajectory of each fish using the nearest-neighbor method and resampled these time series with a 4-second resolution in order to synchronize the trajectories of all fish in each trial and pond. Using these synchronized trajectories, the polarization order parameter is defined by~\cite{kolpas2007coarse, tunstrom2013collective}
\begin{equation}\label{eq:polarization}
    O_p = \frac{1}{N} \Bigg|\sum_{i=1}^N \vec{u}_i\Bigg|\,,
\end{equation}
where $\vec{u}_i$ is the unit direction vector of the $i$-th fish in a given time for a particular trial and pond with $N$ tracked fish. The values of $O_p$ are restricted to the interval $[0,1]$ and $O_p\approx 1$ represents a configuration where fish are strongly aligned, while $O_p\approx 0$ indicates a situation with no alignment on average. In its turn, the rotation order parameter is calculated from~\cite{kolpas2007coarse, tunstrom2013collective}
\begin{equation}\label{eq:rotation}
    O_r = \frac{1}{N} \Bigg|\sum_{i=1}^N \vec{u}_i\times \vec{r}_i\Bigg|\,,
\end{equation}
where $\vec{r}_i$ is a unit vector pointing from school's center of mass towards the position of the $i$-th fish. This quantity represents the normalized average values of the fish angular momentum and it is constrained to interval $[0,1]$. The lower bound of $O_r$ represents a situation with no rotation, while the upper bound indicates a strong rotating configuration. The results of Fig.~\ref{figure3} were obtained considering only trials with five or more fish; however, our conclusions remain the same when considering the entire data set.

\subsection*{Interaction networks}

We use the same synchronized trajectories defined in the previous section to create the interaction networks. For a given trial and pond with five or more fish, we select all positions at a particular time $t$ and create the Voronoi diagram. This diagram is a partitioning of the fish positions plane into convex polygons so that each polygon contains the position of one fish, and every point within the polygon is closer to this particular fish than to any other fish (see black lines of Fig.~\ref{figure4}A for an example). Using the Voronoi diagram at time $t$, we create a graph $g_t = (\nu,\varepsilon)$ where vertices $\nu$ correspond to fish and the set of edges $\varepsilon$ connects all pairs of individuals that are neighbors in the Voronoi tessellation (see orange lines of Fig.~\ref{figure4}A for an example). Finally, we combine these graphs $g_t$ using non-overlapping time intervals of length $\Delta t$ to create a weighted network $G_\tau = (V,E,W)$ where $\tau$ stands for the central point in the time interval, vertices $V$ represent fish, the set of edges $E$ connect all pairs of individuals that were neighbors at least once within the time interval, and $W$ is the set of edge weights representing the fraction of time a pair of individuals remained neighbors during the time interval divided by the number of fish minus one. The results of Fig.~\ref{figure3} were obtained for $\Delta t=10$~min; however, sensitivity analysis reveals similar results when considering other time intervals (see Fig.~S7). Furthermore, we consider a network to represent the under shade category when more than 10\% of fish positions are localized under the shade structures within the time interval (sensitivity analysis reveals similar results when considering different thresholds, as shown in Fig.~S8). 

\subsection*{Network measures}

\noindent {\it Average weighted degree.} The weighted degree $\kappa$ of a vertex is the sum of all weights of its incident edges~\cite{newman2010networks}. The average weighted degree $k$ of a network is the average value of the weighted degree of all network vertices.
\vspace{0.25em} 

\noindent {\it Average weighted transitivity.} Transitivity measures the extent to which vertices in a graph cluster together. For simple and undirected graphs, transitivity (or the global clustering coefficient) is defined as the ratio between the number of closed triplets and the total number of triplets (open and closed)~\cite{newman2010networks}. This measure quantifies the tendency of the nearest neighbors of a particular node to be neighbors themselves. There are several generalizations of transitivity for weighted networks~\cite{saramaki2007generalizations}, and we have used Onnela~\textit{et al.}'s definition~\cite{onnela2005intensity} that replaces the number of closed triplets in the unweighted definition by the normalized sum of triangle intensities (where intensity refers to the geometric mean of edge weights), as implemented in the Python package NetworkX~\cite{hagberg2008exploring}.

\bibliography{references.bib}

\section*{Data availability}

The data which support this manuscript are available at [Faulkner, J.~D., Acre, M.~R. \& Brey, M.~K., 2022, Silver Carp (\textit{Hypophthalmichthys molitrix}) locations in earthen ponds with overhead structure: U.S. Geological Survey Data Release, \url{https://doi.org/10.5066/P9XURDHS}].

\section*{Author contributions statement}

H.V.R, M.R.A., J.D.F., L.R.C., K.M.L., J.J.W., M.K.B., C.M.W., and R.D.C. designed research, performed research, analyzed data, and wrote the paper.

\section*{Acknowledgements}

The authors thank Andy Mueller, Cody Slaugh, Mark Roth, Steve Shier, and Taylor Tidwell for assistance with fish tagging and data collection. The authors would also like to thank Matt Sholtis, Nick Swyers, and Ty Hatton for processing the telemetry data. This study was completed under the auspices of U.S. Geological Survey Institutional Animal Care and Use Committee at the Columbia Environmental Research Center (IACUC number: AEH-18-CERC-01). Funding sources were the U.S. Geological Survey Ecosystems Mission Area Invasive Species Program. Any use of trade, firm, or product names is for descriptive purposes only and does not imply endorsement by the U.S. Government. H.V.R. and L.R.C. acknowledge the support of the Coordena\c{c}\~ao de Aperfei\c{c}oamento de Pessoal de N\'ivel Superior (CAPES) and the Conselho Nacional de Desenvolvimento Cient\'ifico e Tecnol\'ogico (CNPq -- Grant 303533/2021-8). There is no conflict of interest declared in this article.

\clearpage
\includepdf[pages=1-9,pagecommand={\thispagestyle{empty}}]{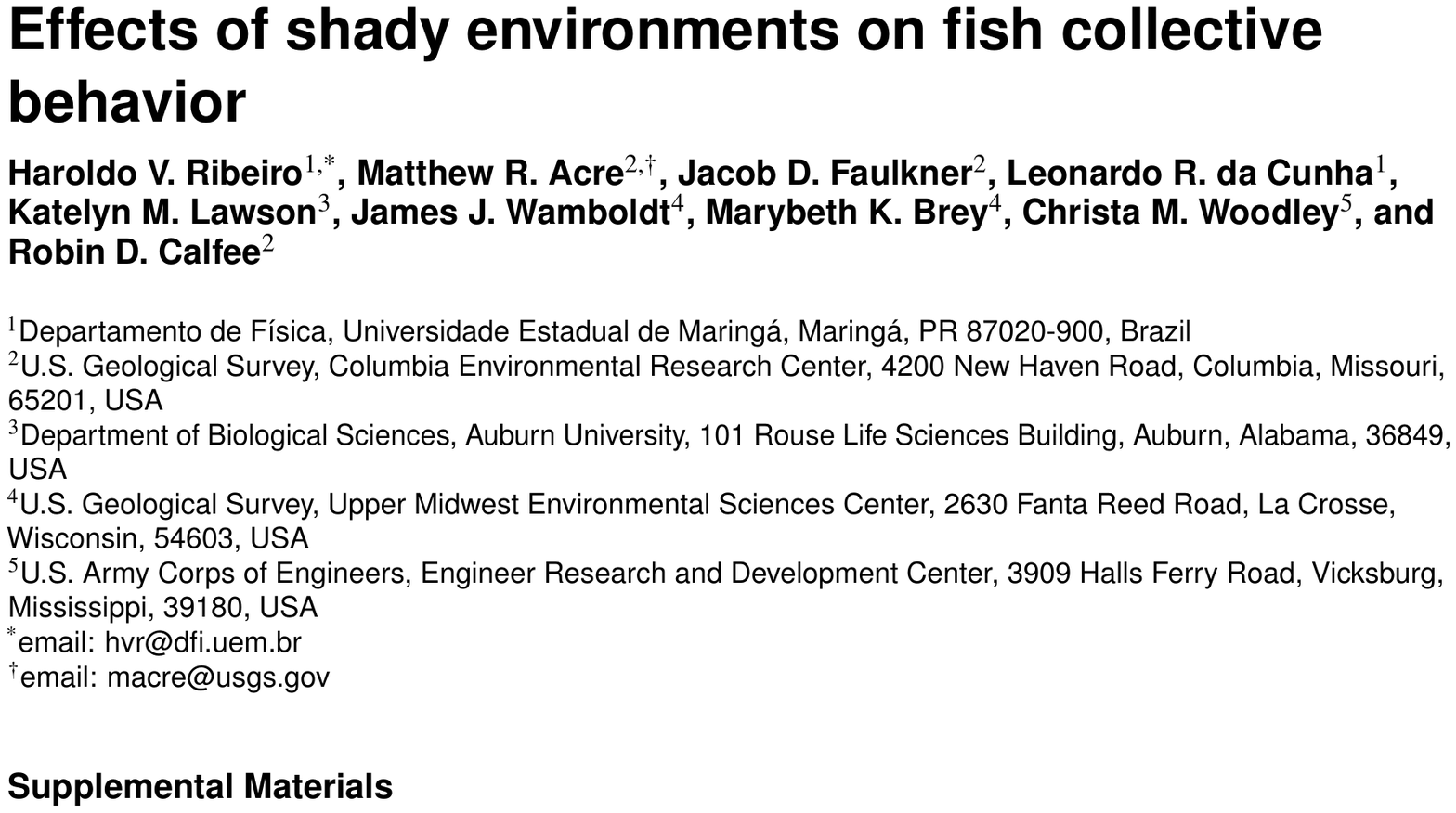}

\end{document}